# Transparent dielectric metasurfaces for mode modulation and spatial multiplexing

Sergey Kruk[1,*], Filipe Ferreira[2], Naoise Mac-Suibhne[1], Christos Tsekrekos[2], Ivan Kravchenko[3], Andrew Ellis[2], Dragomir Neshev[1], Sergey Turitsyn[2], and Yuri Kivshar[1]

[1]*Nonlinear Physics Center, Australian National University, Canberra ACT 2601, Australia*
[2]*Aston Institute of Photonic Technologies, Aston University, Birmingham B4 7ET, UK*
[3]*Center for Nanophase Materials Sciences, Oak Ridge National Laboratory, TN 37831, USA*
*Corresponding author: sergey.kruk@anu.edu.au

Expanding the use of physical degrees of freedom to employ spatial multiplexing of data in optical communication is considered the most disruptive and effective solution to meet the capacity demand of the growing information traffic. Development of space-division-multiplexing methods stimulated research on spatial modulation, detection and processing of data, attracting interest from various fields of science. A passive all-dielectric metasurface with near-unity transmission is used to engineer spatial mode profiles, potentially of arbitrary complexity. The broadband response of the metasurface covers S, C, and L bands of fibre communications. Unlike conventional phase plates, the metasurface allows for both phase and polarization conversion, providing full flexibility for mode engineering. We employ the metasurface for both mode modulation and mode multiplexing in free-space optical communication, and demonstrate that it is capable of mode multiplexing with an extinction ratio in excess of 20 dB over the whole C-band with negligible penalty even for 100 Gb/s DP-QPSK signals. These results merge two seemingly different fields, optical communication and metamaterials, and suggest a novel approach for ultimate miniaturisation of mode multiplexers and advanced LiFi technologies.

Optical communication is rapidly approaching the limits of its current technologies to cope with the fast growing demand on capacity from the existing and emerging applications and broadband services. Over the past 40 years the amount of information transferred over optical communication networks has been increasing by approximately an order of magnitude every four years and by now has long surpassed the Zettabyte per year [1,2]. The key to increase the information capacity was light modulation and multiplexing across four physical dimensions: time, wavelength, polarization and quadratures (amplitude and phase modulation). However, the achievable data transmission rates of the existing single-mode fiber transmission technologies are reaching their limits, due to the nonlinear propagation effects, saturating available capacity [3].

Capacity is proportional to the number of independent communication channels. Therefore, a promising and feasible way to overcome the foreseeable capacity crunch is to look beyond the four physical dimensions used for light modulation and multiplexing. Electromagnetic fields can be discriminated across a fifth dimension: space. The spatial dimension can be exploited by sending information over several parallel spatial paths. This concept is widely used in other communication areas, such as printed circuit boards, Ethernet cables, multi-antenna techniques in cellular wireless systems, etc. Similar techniques have been employed for free-space data transmission based on orbital angular momentum multiplexing [4-7]. Recently, optical communication research has focused on fibers that allow for *spatial division multiplexing* (SDM) in terms of mode division multiplexing [1,2]. A number of technical challenges have to be overcome before SDM will become a technology of choice. In particular, new solutions for control and manipulation of spatial modes are of high practical interest.

Here, we demonstrate a way towards miniaturisation and integration of mode multiplexers and demultiplexers by using highly transparent all-dielectric metasurfaces. Metasurfaces are ultra-thin patterned structures that emerged as a type of *planar metadevices* [8] capable of reshaping and controlling the wavefront of incident light. Various demonstrations of functional optical elements were done recently with all-dielectric metasurfaces [9]. Here we design a metasurface that engineers LP fiber modes – a standard mode basis for fibre communications. We experimentally demonstrate a highly-efficient simultaneous conversion, and thus multiplexing, of LP01 modes into LP11 and LP21 modes in a free-space optical communication configuration. The approach can be readily extended to higher-order modes. With this, we suggest a novel way for miniaturised mode conversion and multiplexing in optical SDM systems. Besides multiplexing, we demonstrate mode modulation with metasurfaces that enable to encode information into spatial domain. The ability to both multiplex and modulate information by using spatial modes fully opens the potential of the spatial dimension as a degree of freedom in next generation optical transmission systems for both free space and fibre communications.

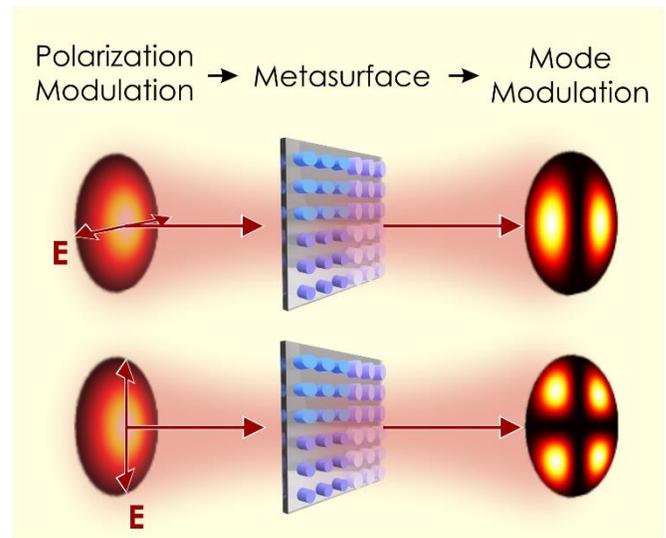

FIG. 1. Concept of the mode multiplexing and mode modulation in spatial domain with metasurfaces. An all-dielectric metasurface of sub-micron thickness and near-unity transmission efficiency converts incoming Gaussian LP01 modes into LP11 and LP21 modes, enabling mode multiplexing or mode modulation.

## Design of transparent metasurfaces

Metasurfaces are composed of subwavelength elements that are distributed spatially across a surface. Due to resonant scattering, each element can alter the phase, amplitude, and polarization of the incoming light. Many designs and functionalities of metasurfaces suggested so far are based largely on plasmonic planar structures [10]; however, most of these metasurfaces demonstrate low efficiencies in transmission due to losses in their metallic components. In contrast, all-dielectric metasurfaces are based on lattices of subwavelength resonant dielectric elements [11] that allow to avoid absorption losses, enhancing substantially the overall efficiency of planar meta-optics [12-17] making all-dielectric metasurfaces a promising platform for diverse applications in optical communications.

We employ the approach suggested earlier in Refs. [14,15] based on the generalized Huygens' principle and design a passive all-dielectric metasurface of sub-micron thickness and high transmission efficiency. Generalized Huygens' metasurfaces were demonstrated to act as efficient transparent phase holograms of high complexity [15], and therefore they can be readily designed and fabricated as converters between arbitrary spatial modes. We demonstrate a single metasurface that acts as two independent phase masks for two linear orthogonal polarizations of light. In conjunction with polarization modulator the metasurface provides a unique capability of ultra-fast mode modulation, thus enabling to encode information into spatial domain [see Fig 1]. We exemplarily demonstrate conversion of a Gaussian LP01 mode into a LP11 or LP21 mode depending on the LP01 polarization state.

To design the metasurface mode converter and multiplexer we use CST Microwave Studio. We employ a set of three different silicon nanopillars arranged into four quadrants (see Fig. 2). All nanopillars have a height of 850 nm and they are placed on a fused silica substrate. Nanopillars in quadrants A have elliptical cross-section with axes 215 x 140 nm. Quadrant B is composed of identical nanopillars rotated by 90 degrees. Quadrants C and D are made of circular nanopillars with radii 180 nm and 150 nm, respectively. To fabricate this dielectric metasurface, we deposit poly-silicon on a silica wafer with low-pressure chemical vapor deposition. Electron beam lithography defines the geometry of the nanopillars and reactive ion etching translates the geometry into silicon [see Fig 2 (a,b) for the fabricated metasurface]. All the quadrants show high optical transmission over the spectral region of the C, S and L optical fibre communication bands. Experimentally measured transmission spectra are shown in Fig. 2(c).

Figure 2(d) shows the experimentally measured phase profiles of the metasurface for two linear orthogonal polarizations at 1550 nm wavelength. The design parameters are optimized to provide π phase delay difference between the arrays C and D. The quadrant A provides the same phase delay as the quadrant C for the horizontal polarization and the same delay as the quadrant D for the vertical polarization. The phase delay of the quadrant B is π-shifted from that of the quadrant A and, correspondingly, it is the same as for quadrant C for the vertical polarization, as well as for quadrant D for the horizontal polarization. Thus, the metasurface serves as a *phase mask* that converts horizontally polarized Gaussian LP01 mode into LP11b mode and vertically polarized Gaussian mode into LP21a mode. Figure 2(e) presents the experimentally observed mode conversion at 1550 nm.

## Mode multiplexing with metasurfaces

In the following, we report our experimental investigation on the ability of our metasurface to simultaneously convert the two incoming orthogonally polarized LP01 signals into two different higher-order LP modes (LP11 and LP21). The performance of the metasurface is compared with that of conventional phase plates [18], able to convert between pair of modes only (e.g. LP01 to LP11 or LP01 to LP21).

The metamaterial mode multiplexer performance was assessed in back-to-back using two conventional phase plates for demodulation in a broadcast and select configuration, as sketch in Fig. 3(a). The metasurface was illuminated with a dual-polarization LP01-mode beam obtained by collimating the output of a standard single-mode fiber (SMF-28). The collimated beam diameter was 0.3mm. To allow the simultaneous measurement of the LP11-mode and LP21-mode, we used a beam splitter at the metasurface, after which a LP21 selective phase plate and a LP11 selective phase plate were used in the transmission and reflection path, respectively. Finally, the two demodulated modes were coupled back into a SMF-28.

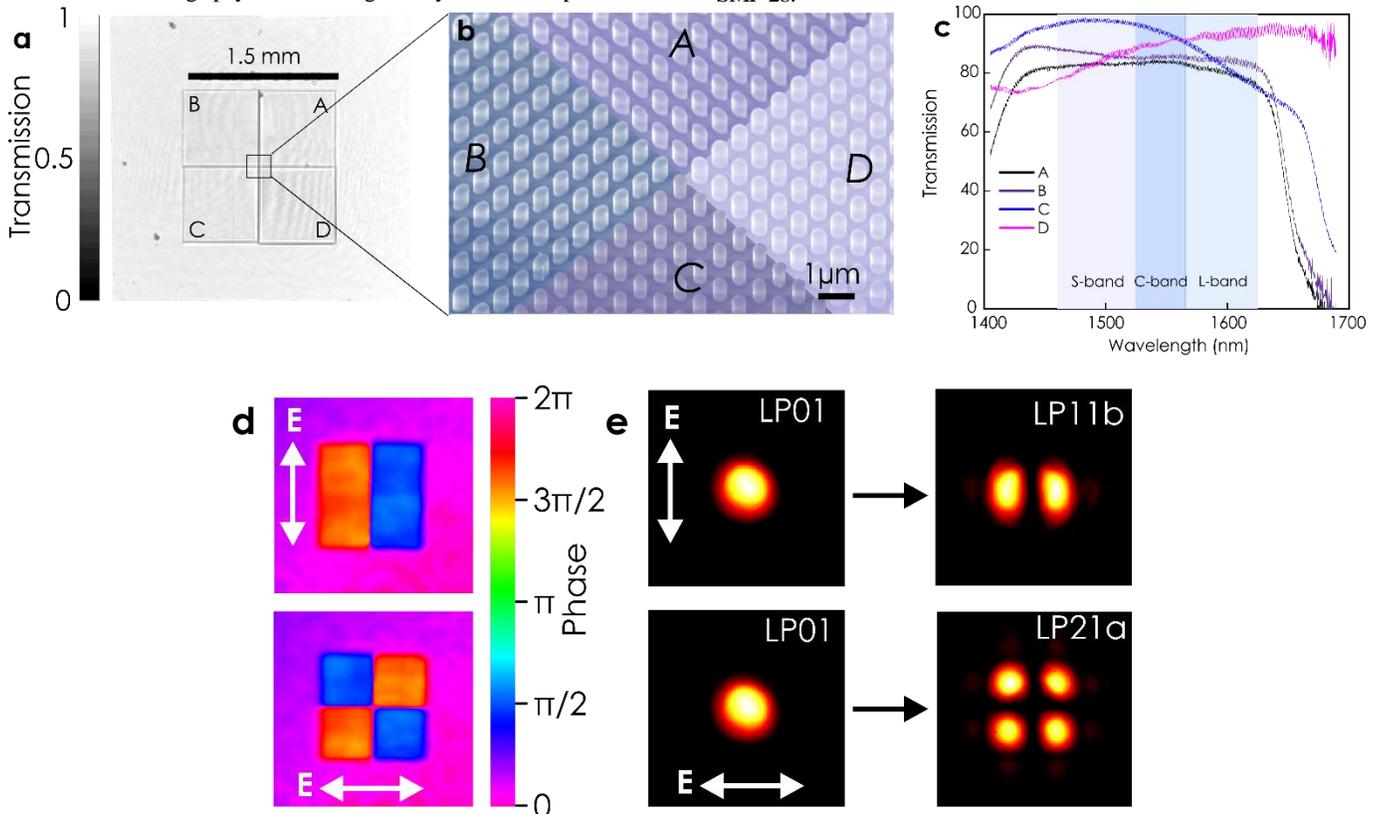

FIG. 2. (a) Image of the metasurface multiplexer in an optical transmission microscope. Illumination spectral range is 1530-1560 nm. (b) SEM image of a central part of the metasurface. (c) Transmission spectra of the four quadrants of the metasurface. (d) Experimentally measured phase accumulation across the metasurface for vertically and horizontally polarized light, respectively. (e) Experimentally observed mode conversion: (top) vertically polarised LP01→LP11b and (bottom) horizontally polarised LP01→LP21a

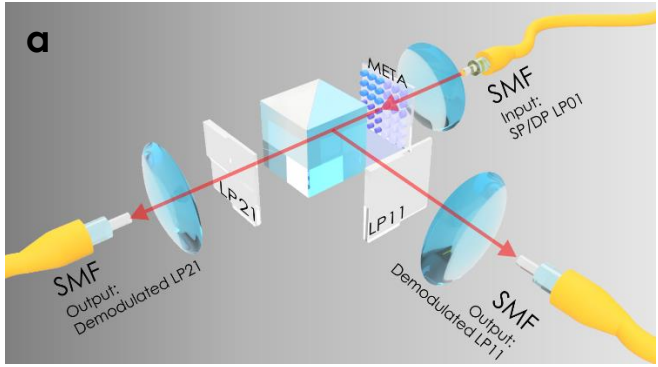

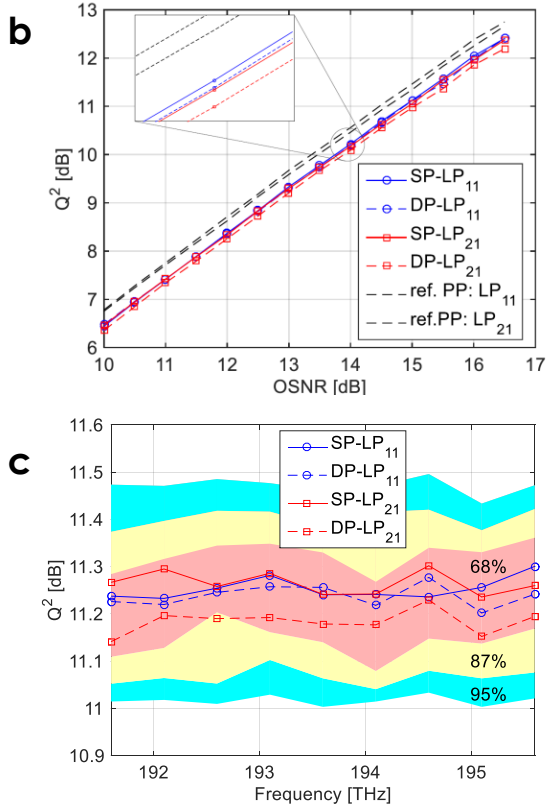

FIG 3. (a) Experimental apparatus for the assessment of the metasurface mode multiplexer performance. (b) Q-factor as a function of OSNR for one channel at 193.6 THz using a 100 Gb/s dual-polarization QPSK lab transponder (results averaged over 20 repetitions per OSNR value). SP and DP stand for single and dual polarization; PP is a phase plate as a reference device. (c) Q-factor as a function of channel frequency over the C-band with a mechanical alignment for 193.6 THz. The colored lines show the Q-factor averaged over 20 repetitions per OSNR value. The colored surfaces identify the area containing the Q-factor values with a probability of 68%, 87% and 95%. SP – single polarization; DP - dual polarization; PP – phase plate as a reference device for SP operation.

The metasurface was tested for mode conversion extinction ratio using continuous wave (CW) signals in a back-to-back configuration and using conventional phase plates (in a broadcast and select arrangement) for mode demodulation. The extinction ratio was defined as the ratio between the power of the desired mode (LP21 or LP11) and the power of the undesired mode (LP11 or LP21). The power was measured at the two SMF outputs following the phase plates using in a broadcast and select arrangement to demux the transformed optical beam at the metasurface output. The extinction ratio was measured to be higher than 22 dB, similar to that of the two conventional phase plates used for the two modes demodulation (25 dB).

After measuring the extinction ratio, we measured the signal quality factor (Q-factor) for a relevant range of optical-signal-to-noise-ratios, and for the central wavelength of the C-band used for optical communications. The Q-factor is measured for single-polarization and dual-polarization signals. Figure 3(b) shows the Q-factor when using a 100 Gbit/s dual-polarization QPSK lab transponder [16], averaged over 20 repetitions. Figure 3(b) shows that the metasurface can convert polarization-multiplexed signals into mode-multiplexed signals with the same performance as it converts single-polarization signals into single higher-order LP modes. Further results shown a maximum 0.2 dB fluctuation between 20 different repetitions. Finally, the performance is compared to that of single-polarization conversion using conventional phase plates. The required OSNR for a given Q-factor is 0.4 dB higher than for the conventional phase plates given the metasurface slightly lower extinction ratio (3 dB lower).

Finally, we measure the Q-factor for wavelengths across the whole C-band used for optical communications for an OSNR of 15 dB. Figure 3(c) shows the Q-factor when using a 100 Gbit/s dual-polarization QPSK lab transponder across the C-band. From the figure, it can be seen that the Q-factor fluctuation over the C-band and over 20 repetitions for each wavelength, is typically lower than 0.5 dB. These results show that the metasurface is able to operate over the entire telecoms grid.

We should also mention that these results can be useful for visible light communications (or LiFi) that exploits the ability to modulate visible LEDs at high speed for wireless communication applications [20], and transmit parallel data streams scaling up the data capacity by spatial multiplexing and high-level modulation schemes [21].

In summary, we have suggested that metadevice multiplexers made of all-dielectric metasurfaces readily allow for engineering mode profiles of arbitrary complexity, including Eisenbud–Wigner–Smith states [22] and orbital angular momentum modes [5]. We have demonstrated, for the first time to our knowledge, that highly transparent all-dielectric resonant metasurfaces can be employed for engineering the mode profiles with high efficiency over a broadband spectral range covering telecommunication S, C, and L bands. We have realized experimentally the conversion of LP01 modes into LP11 and LP21 modes for free-space optical communication. In this way, we have confirmed that a single metasurface is capable of mode-multiplexing with an extinction ratio in excess of 20 dB over the C-band with negligible penalty even for 100 Gb/s DP-QPSK signals. The metasurfaces introduce no performance degradation except for an excess loss due to reflections which can be minimized by applying an antireflective coating. Finally, we have shown that this metadevice is capable of mode-multiplexing operation while previously one required two phase plates with increase free-space arrangement requirements and bulkiness, suggesting a novel approach for ultimate miniaturisation of mode multiplexers and advanced LiFi technologies. It was predicted in [23] that the extension of the current trends in global communications to ~2030 and 2035 would, respectively, require systems with an increased number of parallel spatial paths. Therefore, spatial parallelism and development of SDMxWDM systems is an unavoidable requirement. It was also stressed that a substantial research progress will have to take place across multiple areas, from system architectures to digital signal processing to integrated arrayed device designs in order to avoid an otherwise imminent optical networks capacity crunch. We believe that metasurfaces may offer new designs for the fundamental building blocks of future systems that will represent radical advance in technology.

**Acknowledgment**. The authors acknowledge Katie Chong for the initial work on metasurface masks and Richard Winfield (Tyndall National Institute) for providing the reference phase plates. Fabrication was conducted at the Center for Nanophase Materials Sciences, which is a DOE Office of Science User Facility. This work has been supported by the Australian Research Council, the Liverhulme Visiting Professorship program, the Engineering and Physical Science Research Council (EPSRC) under grant number EP/L000091/1 (PEACE), and EC 7th Framework Programme through Grants 627545 (SOLAS), 659950 (INVENTION) and 654809 (HSPACE).